\documentclass[oneside,english]{elsart}
\usepackage{pslatex}
\usepackage[T1]{fontenc}
\usepackage[latin1]{inputenc}
\usepackage{graphicx}

\makeatletter

\def\Re{{\rm I}\! {\rm R}}
\usepackage{ae}

\usepackage{babel}
\makeatother
\begin{document}
\begin{frontmatter}

\title{Dynamic coordinated control laws in multiple agent models }

\author{David S. Morgan and Ira B. Schwartz}

\address{US Naval Research Laboratory, Plasma Physics Division, Code 6792,
Nonlinear Systems Dynamics Section, Washington, DC 20375}

\thanks{IBS is supported by the Office of Naval Research, DSM is a National
Research Council postdoctoral fellow. }

\begin{abstract}
We present an active control scheme of a kinetic model of swarming.
It has been shown previously that the global control scheme for the
model, presented in \cite{JK04}, gives rise to spontaneous collective
organization of agents into a unified coherent swarm, via a long-range
attractive and short-range repulsive potential. We extend these results
by presenting control laws whereby a single swarm is broken into independently
functioning subswarm clusters. The transition between one coordinated
swarm and multiple clustered subswarms is managed simply with a homotopy
parameter. Additionally, we present as an alternate formulation, a
local control law for the same model, which implements dynamic barrier
avoidance behavior, and in which swarm coherence emerges spontaneously.
\end{abstract}
\end{frontmatter}

\section{Introduction}

Multiple agent models are comprised of a multitude of simple autonomous
vehicles, which are loosely coupled via communication. It is anticipated
that such systems will play a key role in future deployments, as the
drive to miniaturize electronic devices results in smaller and more
capable self-mobile machines with limited decision making abilities.
Thus, one of the main research areas of interest is the dynamic pattern
formation and control of a large number of agents \cite{Bonabeua99}.
In particular, \textit{\emph{given a specific dynamical system composed
of a large number of individual vehicles, each with specified limited
decision-making and communication abilities, a vital question is under
what conditions large-scale aggregate dynamics may be controlled to
form coherent structures, or patterns. An example from electronics
is}} a concept paper \cite{JK01} which shows that complex patterns
can arise from a large array of micro actuators interconnected to
mimic a finite difference approximation of standard reaction diffusion
partial differential equations (PDE). However, it is a static theory
based on quite standard pattern formation theories from reaction-diffusion
which assumes pure local coupling.

In contrast, many biological examples of coherent dynamical motion
(swarming) exist in nature. Populations such as bees, locusts, and
wolves often move in coordinated but localized efforts toward a particular
target. In addition many more examples abound of populations of individuals
that move according to local rules, and whose aggregate dynamics achieve
an overall large-scale complex pattern or state. Bacterial colonies,
which evolve in part via chemotactic response, are such an example.
The mathematical biology community has been exploring models for animal
swarms, and this work pinpoints some of the difficulties (see the
survey paper \cite{EK01}). Traditional models for biology populations
involve local PDE for the population density \cite{Murray}. Edelstein-Keshet
\emph{et al} \cite{EKWG98} recently considered such a model in one
space dimension for African migratory locusts. These insects have
a gregarious phase in which swarms of individuals can travel for days
over thousands of miles. Evidence exists that the swarms remain cohesive
even in the absence of a nutrient gradient. The analysis of \cite{EKWG98}
shows that such cohesive swarms cannot be described by traveling wave
solutions of their one dimensional advection-diffusion model. More
recently, Mogilner and Edelstein-Keshet consider nonlocal interactions,
in which the drift velocity of the population is determined by a convolution
operator with the entire population \cite{MEK99}. These models, resulting
in integro-differential equations, do sometimes produce coherent band-like
structure. Earlier work by Edelstein-Keshet and Watmough \cite{WEK95}
on army ant swarms, considers a one dimensional model and shows the
existence of traveling wave solutions for the leading edge of the
pack, but they do not consider band-like solutions that would describe
something like a locust swarm. These particular examples involve one-dimensional
models and simulations. In summary, most studies of biological swarming
involve models from continuum theory, many of which are based on some
form of local communication, which are modeled by way of interactions
or couplings.

The statistical physics community has recently tried to understand
similar problems in situations where the number of individuals are
very large. Statistical information derived for large numbers is less
relevant to sensor applications involving smaller numbers of individuals.
However, the connection between the discrete and the continuous is
an important problem that is well-studied in this field. The particle
approach involves starting with simple rules of motion, involving
combinations of biased random walks, sampling of motions and positions
of nearby neighbors, with some governing strategy designed to mimic
core components of animal interactions. For example, Schweitzer et
al \cite{SET01} consider a theory of canonical-dissipative systems
and the energetic conditions for swarming. Daniel Gr\"{u}nbaum \cite{G99}
has derived advection diffusion equations for internal state-mediated
biased random walks. Mogilner and Edelstein-Keshet \cite{MEK96} consider
both continuum and cellular automata models for populations of self-aligning
objects. St\"{o}cker \cite{Stocker} considers a hexagonally based
cellular automata model for tuna school formation. These are just
a few examples. In all cases, the local rules are precisely defined
and aggregate motion can be observed in numerical simulations. 

As an alternative to understanding coherent swarm structures that
use finite models (non-continuum theories), a recent body of work
considers general particle-based models for self-propelled organisms
(see for example \cite{Albano,BDG97,CBV99,VCFH99}). Collective motion
and swarming is observed along with interesting aspects of dynamic
phase transitions, including crystalline like motion, liquid, solid,
and gas-like states. Toner and Tu \cite{TT95,TT98,Tu00} use renormalization
group ideas to study flocking motion in a particle-based model. Some
of this work parallels classical statistical theory of transport which
derives hydrodynamic equations from local interaction models \cite{Irving:1950,Thompson:1972,Mazo:1967}.
The approach considered by Chang, \emph{et al.} \cite{ChangShaddenMarsden}
considers agents in a scalar potential field and utilizes gyroscopic
and braking forces.

In most cases presented, the agents are self-propelled and the nature
of the coupling or communication imposes a given pattern. Here we
consider similar aspects, but with the idea of controlling the communication
to form patterns. In this article we consider kinetic models in which,
depending on the control law used, the self-propelled agents communicate,
either locally within a specified radius about each agent, or globally
with every other agent in the swarm. Under appropriate choices of
controlling {}``potentials'', coherent motion of agents is observed.
In general, the models considered are based on controls which involve
long range attraction and short range repulsion, similar to the ideas
in \cite{zohdi03}. However, in \cite{zohdi03}, the computational
approach to obstacle avoidance, achieved by forming clustered groups
from a single coherent swarm, is to use genetic algorithms, which
contain a number of restrictive rules. This violates the assumption
of creating a swarm with limited computational ability. 

In the work presented here, we consider the problem of dynamically
deforming a single large and coherent swarm into a collection of subswarm
clusters under simple control modifications. A cluster is a subset
of the original swarm which functions independently as a coherent
swarm, and which, when fully formed, does not interact with agents
that are not members of the cluster. A primary goal of this work is
to generate simple algorithmic controls for obstacle avoidance, and
we consider two methods to achieve this. We also consider multiple
approaches to guiding a swarm, by dynamically steering leader agent(s),
and by \emph{a priori} fixing a target to which all agents are attracted. 

We formulate the first control problem using homotopy, or continuation,
theory \cite{Allgower80,Rheinboldt00}. The homotopy parameter controls
the communication coupling, selecting between local and global communication,
and may simultaneously be used to modify other characteristics of
the control law. Such a control law allows one to use a single parameter
to switch from a single coherent swarm state (global coupling) to
a multiple cluster state (local coupling) and back again. Swarm coherence
and inter-agent collision avoidance is achieved with this control
law via a long-range attractive/short-range repulsive potential, and
swarm navigation is implemented via group-averaged motion and leader-following
controls.

We also consider an alternate formulation using only local coupling
between agents, in which a convex barrier is detected and avoided,
and where the barrier location is not \emph{a priori} known. This
approach to obstacle avoidance is similar in nature to that discussed
in \cite{ChangShaddenMarsden}. In addition, swarm navigation is achieved
by introducing terms in the control law so that all agents seek a
common target. Whereas clustering with the homotopy control law is
due to an attractive potential to other agents, clustering appears
to arise naturally with this control law, as agents interact while
they seek out a common target.

The layout of the paper is as follows: In section 2 we introduce the
kinetic model presented in \cite{JK04}, and discuss some important
properties of its global control law. In Section 3 we introduce a
modified control law implementing local coupling. A homotopy control
law is presented in Section 4. In Section 5 we present the alternative
control law with an alternate approach to target seeking behavior
and barrier avoidance, and we conclude with a discussion.

\section{\label{sec:UAV-model-and}Multi-agent kinetic model and properties}

The ideas we present apply to a large class of systems. Consider a
continuous dynamical system $\frac{d\mathbf{z}}{dt}=\mathbf{F}(\mathbf{z}(t))$
arising from an autonomous vector field $\mathbf{F}$, where $t\in\Re$
and $\mathbf{z}\in\Re^{n}$, describing the equations of motion. Associated
to this dynamical system is the system governing trajectories, in
which all orbits have unit velocities,\begin{equation}
\frac{d\mathbf{r}}{dt}=\mathbf{G}(\mathbf{r}(t))\equiv\frac{\mathbf{F}(\mathbf{r})}{\left\Vert \mathbf{F}(\mathbf{r})\right\Vert }.\label{eq:traj_equation}\end{equation}
Consider a (nontrivial) trajectory $\mathbf{r}(t)$ of (\ref{eq:traj_equation}),
its associated unit tangent vector $\mathbf{x}=\mathbf{G}(\mathbf{r}(t))$
defined for all $t\in\Re$, and the positively oriented unit conormal
vectors $\mathbf{y}_{i}(t),\:(i=1,n-1)$. The collection of vectors
$\mathcal{F}=\{\mathbf{x}(t),\mathbf{y}_{i}(t)\:(i=1,\ldots,n-1)\}$
is called the (moving) reference frame associated to $\mathbf{r}$.
Thus, one may recast a continuous dynamical system as a system of
trajectories $\mathbf{r}(t)$ parameterized by arclength, with the
associated moving frame $\mathcal{F}$. The behavior of a system of
trajectories with the associated moving frame is governed by the Frenet-Serret
system of equations.

\subsection{Derivation of Frenet-Serret system of equations}

We derive the Frenet-Serret equations, restricted to the plane, following
the approach of \cite{Jurdjevic}. Consider a differentiable trajectory
$\mathbf{r}(t)$ in $\Re^{2}$, parameterized by arclength, which
represents the motion of an agent over time. A positively oriented
orthonormal frame $\mathbf{x}$ and $\mathbf{y}$ is associated to
$\mathbf{r}(t)$, by taking $\mathbf{x}$ equal to the unit tangent
vector $d\mathbf{r}/dt$, and $\mathbf{y}=\mathbf{x}^{\perp}$ to
be the unit normal vector positively oriented relative to $\mathbf{x}$.
There exists a function $\kappa(t)$, called the curvature of $\mathbf{r}(t)$,
such that\begin{equation}
\frac{d\mathbf{r}}{dt}=\mathbf{x}(t),\quad\frac{d\mathbf{x}}{dt}=\kappa(t)\cdot\mathbf{y}(t).\label{eq:FS_deriv1}\end{equation}
One then obtains the equation governing the unit normal vector \textbf{$\mathbf{y}$}
as follows. Using the right-hand equation of Eq. (\ref{eq:FS_deriv1})
we find that $\mathbf{x}\cdot\mathbf{y}=d\mathbf{x}/dt\cdot\mathbf{y}+\mathbf{x}\cdot d\mathbf{y}/dt=\kappa(t)+\mathbf{x}\cdot d\mathbf{y}/dt=0$,
and thus $d\mathbf{y}/dt=-\kappa(t)\cdot\mathbf{x}$. The moving frame
$\mathbf{x}$ and $\mathbf{y}$ associated with $\mathbf{r}(t)$ can
be expressed by a rotational matrix $R(t)$, which has columns consisting
of coordinates of $\mathbf{x}$ and $\mathbf{y}$ relative to a fixed
orthonormal frame $e_{1}$ and $e_{2}$ in $\Re^{2}$. 

This formulation leads to a natural Lie group setting, but we do not
consider that aspect further in this article. We also note that it
is also possible to derive the Frenet-Serret equations by considering
the problem of steering unit-charge, unit-mass particles in a magnetic
field. For details, consult \cite{JK04} and references therein.

\subsection{Equations of motion for multiple agents.}

We consider a set of $n$ agents, restricted to smooth motions in
the plane, and moving at unit speed. The system of equations modeling
each agent is\begin{eqnarray}
\dot{\mathbf{r}}_{k} & = & \mathbf{x}_{k}\label{eq:two_dim_Frenet-Serret}\\
\dot{\mathbf{x}}_{k} & = & \mathbf{y}_{k}\cdot u_{k}\nonumber \\
\dot{\mathbf{y}}_{k} & = & -\mathbf{x}_{k}\cdot u_{k},\nonumber \end{eqnarray}
for $k=1,\dots,n$. The orientation of an agent is given by the moving
frame $\mathbf{x}$ and $\mathbf{y}$, its trajectory is given by
$\{\mathbf{r}(t)|t\in\Re\}$, and the agents are coupled together
via a scalar curvature control law $u$, which is detailed below. 

The control law $u_{k}$, introduced in \cite{JK04}, is\begin{equation}
u_{k}=\sum_{j\neq k}u_{jk},\label{eq:control_law}\end{equation}
with\begin{equation}
u_{jk}=\left[-\eta\left(\frac{\mathbf{r}_{jk}}{\left|\mathbf{r}_{jk}\right|}\cdot\mathbf{x}_{k}\right)\left(\frac{\mathbf{r}_{jk}}{\left|\mathbf{r}_{jk}\right|}\cdot\mathbf{y}_{k}\right)-f\left(\left|\mathbf{r}_{jk}\right|\right)\left(\frac{\mathbf{r}_{jk}}{\left|\mathbf{r}_{jk}\right|}\cdot\mathbf{y}_{k}\right)+\mu\mathbf{x}_{j}\cdot\mathbf{y}_{k}\right]\label{eq:control_law_atom}\end{equation}
where $\mathbf{r}_{jk}\equiv\mathbf{r}_{k}-\mathbf{r}_{j}$, $f$
is \begin{equation}
f\left(\left|\mathbf{r}_{jk}\right|\right)=\alpha\left[1-\left(\frac{r_{0}}{\left|\mathbf{r}_{jk}\right|}\right)^{2}\right],\label{eq:distance_func}\end{equation}
and $\eta=\eta(|\mathbf{r}|)$, $\mu=\mu(|\mathbf{r}|)$, and $\alpha=\alpha(|\mathbf{r}|)$
are specified functions. We now describe this control law in some
detail. We first note that when $u_{k}<0$ $(u_{k}>0)$, the Frenet
frame will rotate in a clockwise (anticlockwise) fashion, respectively.
In order to simplify the following discussion, we consider the case
of $n=2$, but note that the discussion holds for general $n$. Let
$\mathbf{r}_{1}$, $\mathbf{x}_{1}$ and $\mathbf{y}_{1}$ be the
position and corresponding Frenet frame of one of the agents. We will
examine each of the terms in Eq. (\ref{eq:control_law_atom}) in turn.
The first term, $-\eta(\mathbf{r}_{jk}/|\mathbf{r}_{jk}|\cdot\mathbf{x}_{k})(\mathbf{r}_{jk}/|\mathbf{r}_{jk}|\cdot\mathbf{y}_{k})$
serves to orient the vehicles perpendicular to their common baseline,
$\mathbf{r}_{jk}$. To see this, let $\theta_{\mathbf{x}}$ and $\theta_{\mathbf{y}}$
be the angles the (unit) vectors $\mathbf{x}_{1}$ and \textbf{}$\mathbf{y}_{1}$
make with $\mathbf{r}_{21}/|\mathbf{r}_{21}|=(\mathbf{r}_{1}-\mathbf{r}_{2})/|\mathbf{r}_{1}-\mathbf{r}_{2}|$,
respectively. Then\begin{eqnarray*}
-\eta(\mathbf{r}_{21}/|\mathbf{r}_{21}|\cdot\mathbf{x}_{1})(\mathbf{r}_{21}/|\mathbf{r}_{21}|\cdot\mathbf{y}_{1}) & = & -\eta\cos(\theta_{\mathbf{x}})\cos(\theta_{\mathbf{y}})\\
 & = & -\eta\cos(\theta_{\mathbf{x}})\cos(\theta_{\mathbf{x}}-\frac{\pi}{2}).\end{eqnarray*}
This expression is zero for $\theta_{\mathbf{x}}=\frac{\pi}{2},\frac{3\pi}{2}$,
positive for $0\le\theta_{\mathbf{x}}<\frac{\pi}{2}$ and $\pi\leq\theta_{\mathbf{x}}<\frac{3\pi}{2}$,
and is negative elsewhere. Thus, this term steers the vehicle to the
nearest perpendicular with the baseline $\mathbf{r}_{21}$.

\begin{figure}
\begin{center}\includegraphics[%
  scale=0.75]{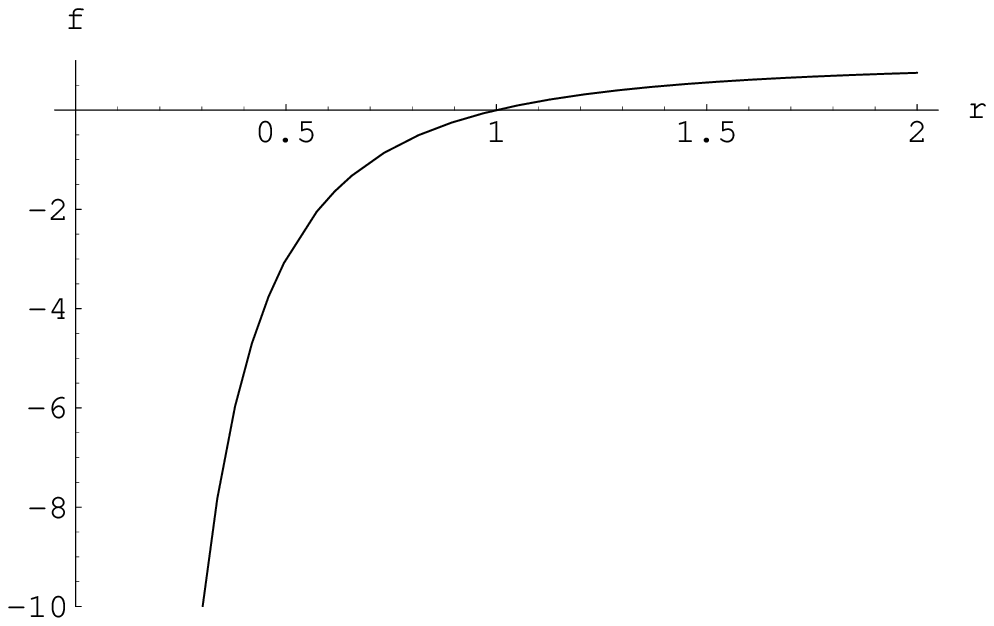}\includegraphics[%
  scale=0.75]{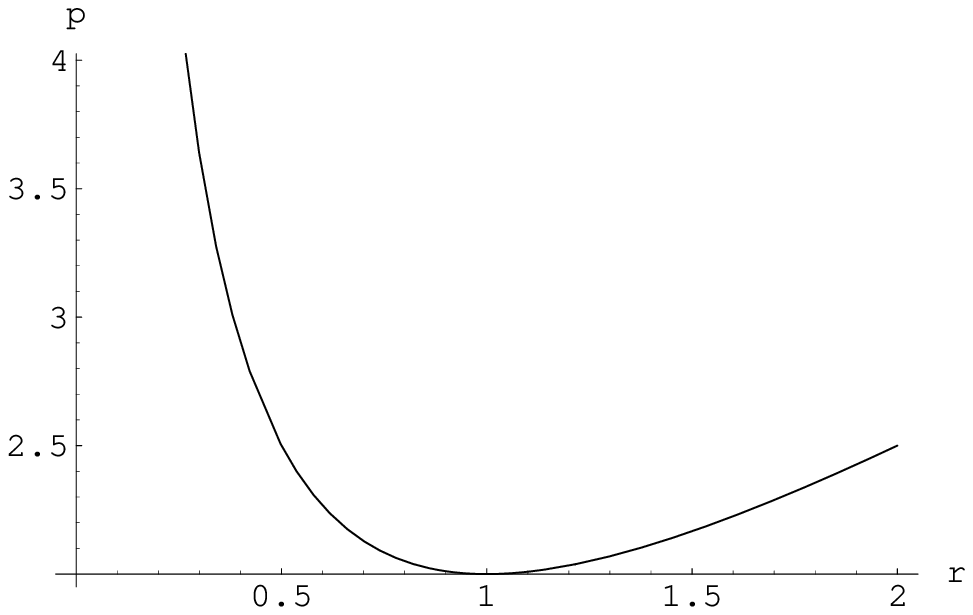}\end{center}

\caption{\label{cap:f_and_potential}This shows the function $f(|r_{jk}|)=\alpha(1-[r_{0}/|r_{jk}|]^{2})$,
and the corresponding (unbounded) potential $p$, for $r_{0}=1$ and
$\alpha=1$. Two agents greater in distance than $r_{0}$ are attracted
to one another, while they are strongly repelled when the distance
is less than $r_{0}$. }
\end{figure}

Inter-agent spacing is controlled via the short-range repulsive/long-range
attractive term,\begin{equation}
-\alpha\left[1-\left(\frac{r_{0}}{\left|\mathbf{r}_{jk}\right|}\right)^{2}\right]\left(\frac{\mathbf{r}_{jk}}{\left|\mathbf{r}_{jk}\right|}\cdot\mathbf{y}_{k}\right).\label{eq:distance_control}\end{equation}
The first factor of (\ref{eq:distance_control}), which arises from
a Leonard-Jones type of potential, is negative if the distance between
two agents is less than $r_{0}$, and positive if the distance is
greater than $r_{0}$. The sign of the second factor is determined
by the orientation of the two agents $\mathbf{r}_{j}$ and $\mathbf{r}_{k}$,
relative to the baseline between them. See figure \ref{cap:f_and_potential},
which shows both the graphs of the potential, and of $f$.%

It is easy to see that the third term, $\mu\mathbf{x}_{j}\cdot\mathbf{y}_{k}$,
serves to drive the vehicles to a common orientation, by rewriting
the dot product in terms of cosines.

The control law (\ref{eq:control_law}) is global (see Fig. \ref{cap:Local-coupling-scenario.}),
meaning that every agent communicates with all other agents in the
swarm. Furthermore, the final orientation (heading) of the swarm is
obtained by group averaged motion, which is in turn determined by
the initial positions and orientations of the agents. We define this
as the globally coupled, group averaged motion law. For the case $n=2$,
rigorous global convergence results have been obtained, by reducing
(\ref{eq:two_dim_Frenet-Serret}) via the symmetry group $\textrm{SE(2)}$
and demonstrating explicitly the existence of a Lyapunov function,
the physical result being that agents will align to the same heading,
perpendicular to their common base-line, and with the appropriate
distance between them. See \cite{JK04} (and \cite{JK02}) for details.
Recently, local convergence results were obtained for the general
case of $n$ agents. See \cite{JK03} for details.

\section{\label{sec:Modified-control-laws.}The leader following control law
utilizing local coupling.}

\begin{figure}
\begin{center}\includegraphics[%
  scale=0.5]{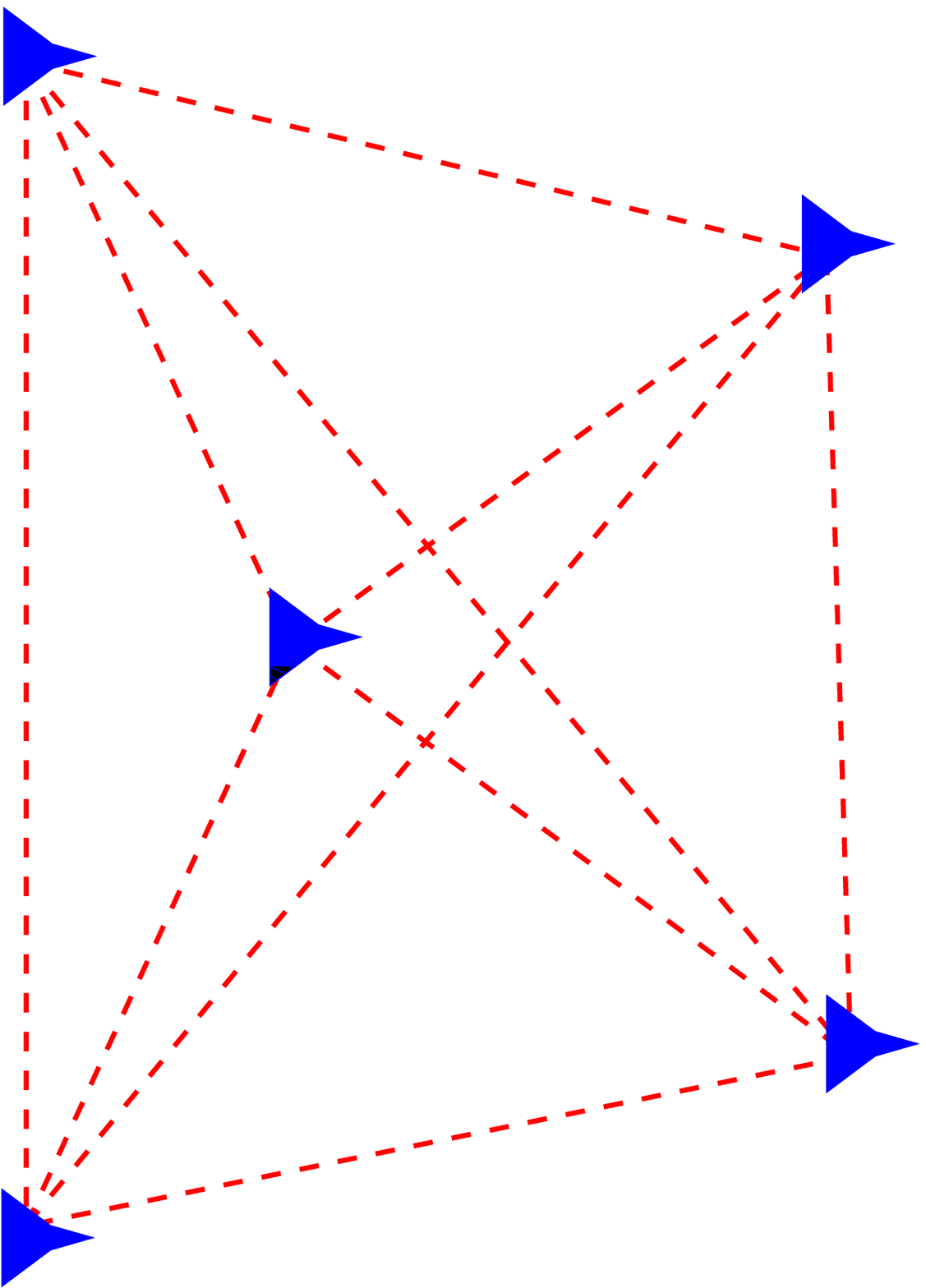}\includegraphics[%
  scale=0.3]{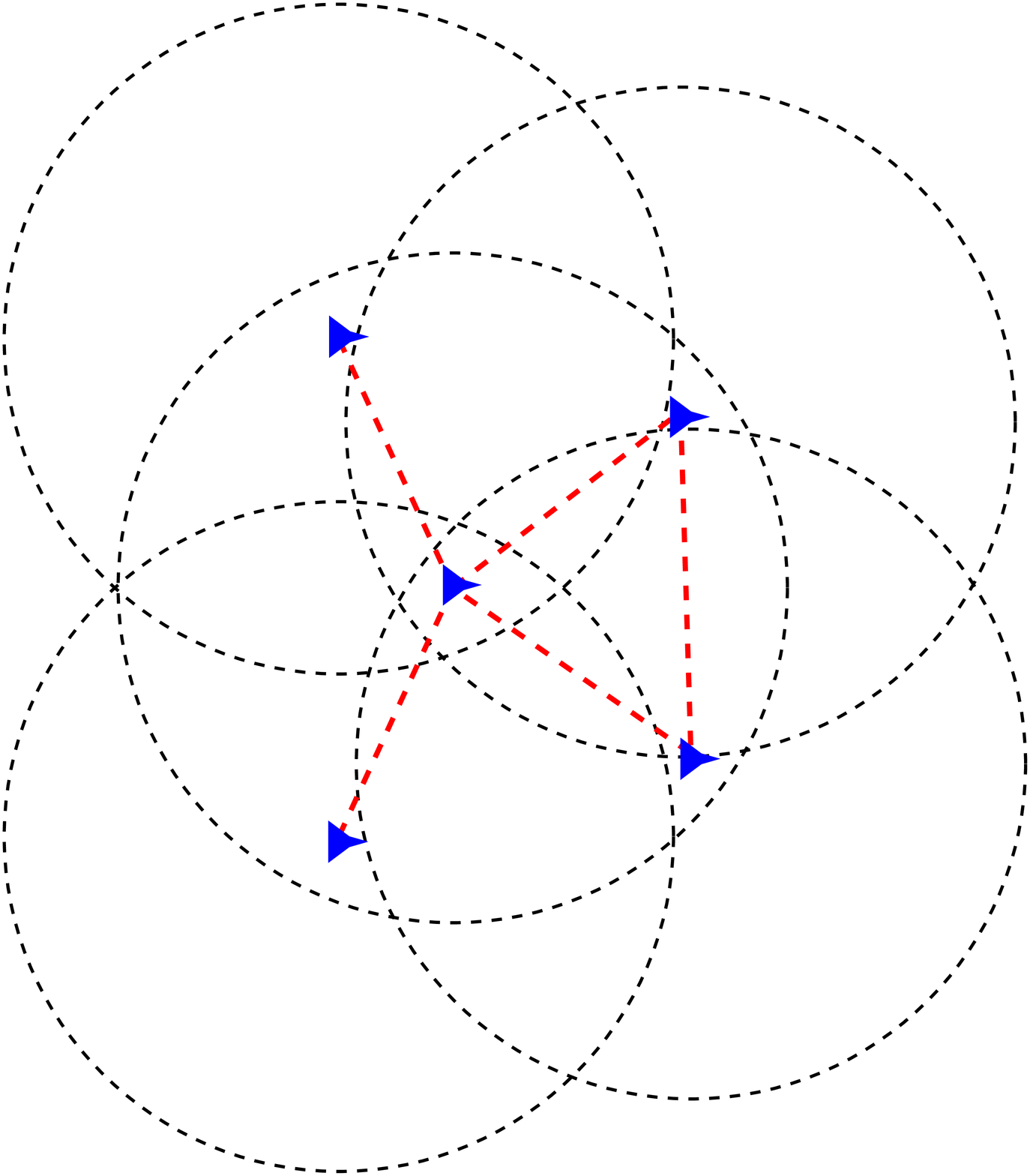}\end{center}

\caption{\label{cap:Local-coupling-scenario.}The left figure shows the global
coupling implemented by the control law (\ref{eq:control_law}). The
right figure shows an example of local coupling scenario implemented
by the control law (\ref{eq:control_law_modified}). Dashed lines
indicate agents which are coupled to one another, while the circles
indicate the maximal communication radius. Note that each agent is
path connected to every other agent. For example, the upper left and
lower right agents are not directly connected, but they are path connected,
since they are both in range of the central agent.}
\end{figure}

The control law (\ref{eq:control_law}) is global; that is, at each
time-step, an agent requires information from all other agents in
the swarm. Global communication is however often not practical. It
is a goal to miniaturize mobile platforms as much as possible, and
so not surprisingly, space constraints limit the power and sensitivity
of on-board sensors and transmitters. Environmental factors, such
as weather effects and local geography can also have detrimental effects
on electromagnetic signals. On the other hand, a local control law
only requires an agent to communicate with some subset of agents in
the swarm, such as their nearest neighbors. Indeed local coupling
is observed in most natural swarms, such as schooling fish and flocking
birds. See Fig. \ref{cap:Local-coupling-scenario.} for a comparison
of the global and local coupling we employ.

We employ local coupling of agents by limiting communication to a
neighborhood of each agent, so that there may be agents that are not
in communication with other agents in the swarm. However, we choose
initial conditions such that each agent is in communication range
of at least one other agent, and such that all agents are `path connected'
initially, meaning that any two agents in the swarm are coupled at
least through intermediary agents. See Fig. \ref{cap:Local-coupling-scenario.}
and caption. 

The implementation of the local coupling model is straight-forward.
We simply multiply the control law (\ref{eq:control_law}) with the
cutoff function\begin{equation}
c\left(\left|\mathbf{r}_{jk}\right|,q,w\right)=\left\{ \begin{array}{cc}
1 & \textrm{if $\left|\mathbf{r}_{jk}\right|<w$,}\\
q & \textrm{otherwise.}\end{array}\right.\label{eq:cutfn}\end{equation}
By using a nonzero value for $q$, one obtains a global `cutoff' function,
that is useful for imposing stronger local coupling, while maintaining
weak global coupling (by setting $q\ll1$). For the present discussion
we set $q=0$, so that only when the Euclidean distance between two
agents is less than $w$ will they interact. We thus obtain the modified
law\begin{eqnarray}
u_{k}^{L} & =c\left(\left|\mathbf{r}_{jk}\right|,0,w\right) & \sum_{j\neq k}u_{jk},\label{eq:control_law_modified}\end{eqnarray}
where $u_{jk}$ is defined by Eq. (\ref{eq:control_law_atom}) .

We note that when the distance between all agents is less than $w$,
Eq. (\ref{eq:control_law_modified}) reduces to the original control
law (\ref{eq:control_law}), while if the swarm is split into subswarm
clusters greater than distance $w$ from one another, the subswarms
will evolve independently of one another.

The control law (\ref{eq:control_law}) uses group averaged motion
for swarm control. The asymptotic heading of the swarm is thus determined
by the initial conditions. We wish to control the direction of the
swarm without having to steer each agent individually. We implement
a leader following control which allows one to 'steer' the swarm by
controlling a designated leader agent (or agents). This provides simple
directional control of a swarm, since only the leader agents are steered,
and the nonleader agents, which we define to be follower agents, pursue
leader agents automatically. We note that leader following behavior
can be implemented with either local or global agent coupling. 

The leader following, local control law is obtained by using (\ref{eq:control_law_modified})
for follower agents, modified so there is stronger coupling between
follower and leader agents, \begin{equation}
u_{k}^{follower}=c\left(\left|\mathbf{r}_{l(k)k}\right|,0,w\right)\ell_{c}u_{l(k)k}+\sum_{j\neq k,l(k)}c\left(\left|\mathbf{r}_{jk}\right|,0,w\right)u_{jk},\label{eq:follower_control}\end{equation}
where $\ell_{c}$ is a coupling constant and $l(k)$ is the index
of the leader swarmer closest to the $k^{\textrm{th}}$ follower swarmer,
while for leader agents the control law is simply\begin{equation}
u_{k}^{leader}=s_{k},\label{eq:leader_control}\end{equation}
where $s_{k}$ is an explicit steering program, which can be given
by a trajectory from a dynamical system.

\section{\label{sec:Homotopy-control-law.}Homotopy control law.}

We combine the leader following, local control law given by Eqs. (\ref{eq:follower_control})
and (\ref{eq:leader_control}), introduced in the previous section,
with the group averaged, global control law (\ref{eq:control_law})
of section \ref{sec:UAV-model-and} to obtain a hybrid control law
utilizing a homotopy parameter (defined below), which we hereafter
refer to as a homotopy control law. The introduction of a homotopy
parameter provides a simple mechanism to dynamically switch from one
control law to another.

Let $u_{k}^{G}$ be the global control law (\ref{eq:control_law})
and let $u_{k}^{L}$ be the local control law given by Eqs. (\ref{eq:follower_control})
and (\ref{eq:leader_control}). The homotopy control law \begin{equation}
u_{k}=u_{k}(\lambda),\;0\leq\lambda\leq1\label{eq:homotopy_ctrl_law}\end{equation}
is defined by the properties\[
u_{k}(\lambda=0)=u_{k}^{G},\quad u_{k}(\lambda=1)=u_{k}^{L},\]
along with the property that the control law $u_{k}$ varies smoothly
with $\lambda$.

To implement the homotopy control law (\ref{eq:homotopy_ctrl_law}),
we designate $m$ agents to be leaders, so that there will be $n-m$
follower agents. Additionally, let $l(k)$ be the index associated
to the closest leader (in the Euclidean sense) of the $k^{\textrm{th}}$
follower agent. The homotopy control law for follower agents is \begin{eqnarray}
u_{k}^{follower}(\lambda) & = & c\left(\left|\mathbf{r}_{l(k)k}\right|,1-\lambda,w\right)[(\ell_{c}-1)\lambda+1]u_{l(k)k}+\label{eq:homo_flwr_ctrl}\\
 &  & \sum_{j\neq k,l(k)}c\left(\left|\mathbf{r}_{jk}\right|,1-\lambda,w\right)u_{jk}\nonumber \end{eqnarray}
where $u_{jk}$ is given by (\ref{eq:control_law_atom}) and $\ell_{c}\gg1$
is a constant which couples followers more strongly to their closest
leader agent. This coupling constant was found to be necessary for
the proper swarm-splitting behavior to emerge. The follower agents
must react more strongly to the motion of leader agents, otherwise
some follower agents were observed to escape from a local neighborhood
of the swarm, and would thus no longer interact with it. The homotopy
control law for the leader agent(s) is 

\begin{equation}
u_{k}^{leader}(\lambda)=\sum_{j\neq k}\left[u_{jk}(1-\lambda)+\frac{s_{k}}{n-1}\lambda\right],\label{eq:homo_ldr_ctrl}\end{equation}
where $s_{k}$ is an explicit steering program, which may be supplied
by an external dynamical system. 

When $\lambda=0$, the cutoff function $c\left(\left|\mathbf{r}_{jk}\right|,1,w\right)\equiv1$,
and both the follower control law (\ref{eq:homo_flwr_ctrl}) and leader
control law (\ref{eq:homo_ldr_ctrl}) reduce to the global control
law (\ref{eq:control_law}). When $\lambda=1$ on the other hand,
\[
c\left(\left|\mathbf{r}_{jk}\right|,0,w\right)=\left\{ \begin{array}{cc}
1 & \textrm{if $\left|\mathbf{r}_{jk}\right|<w$,}\\
0 & \textrm{otherwise,}\end{array}\right.\]
and inter-agent coupling is local. The control law for the follower
and leader agents is in this case given by Eqs. (\ref{eq:follower_control})
and (\ref{eq:leader_control}), respectively.

When only one leader is present, the swarm transitions from group
averaged motion to leader following motion, and thus the swarm can
be directed, as $\lambda\rightarrow1$. Note that in this case, the
local coupling plays no significant role in the behavior, other than
perhaps making the swarm more robust to communications difficulties. 

When more than one leader agent is designated, other swarming behaviors
are possible. In this case, the local coupling plays a crucial role.
For example, assume there are two leader agents. As $\lambda\rightarrow1$,
and the two leaders are directed away from one another, the swarm
will effectively split into two subswarm clusters. This is the result
of the local coupling and the fact that follower agents are more strongly
coupled to their \emph{closest} leader agent. As the leader agents
diverge, the follower agents following one leader leave the communications
range of the followers of the other leader, so that they no longer
interact. The sets of equations modeling the two subswarms in this
case decouple.

\subsection{Homotopy control law simulation results}

We present the results of simulations of the modified model using
the homotopy control law (\ref{eq:homotopy_ctrl_law}). We additionally
present the results of simulations using the homotopy control law
with local coupling throughout. That is, we consider the homotopy
control law (\ref{eq:homotopy_ctrl_law}) with the modification\[
u_{k}(\lambda=0)=u_{k}^{GL},\]
where $u_{k}^{GL}$ utilizes group averaged motion, but with local
coupling. For each control law, we ran several simulations, each with
random initial data, as described below. 

\begin{center}%
\begin{figure}
\begin{center}\includegraphics[%
  scale=0.5]{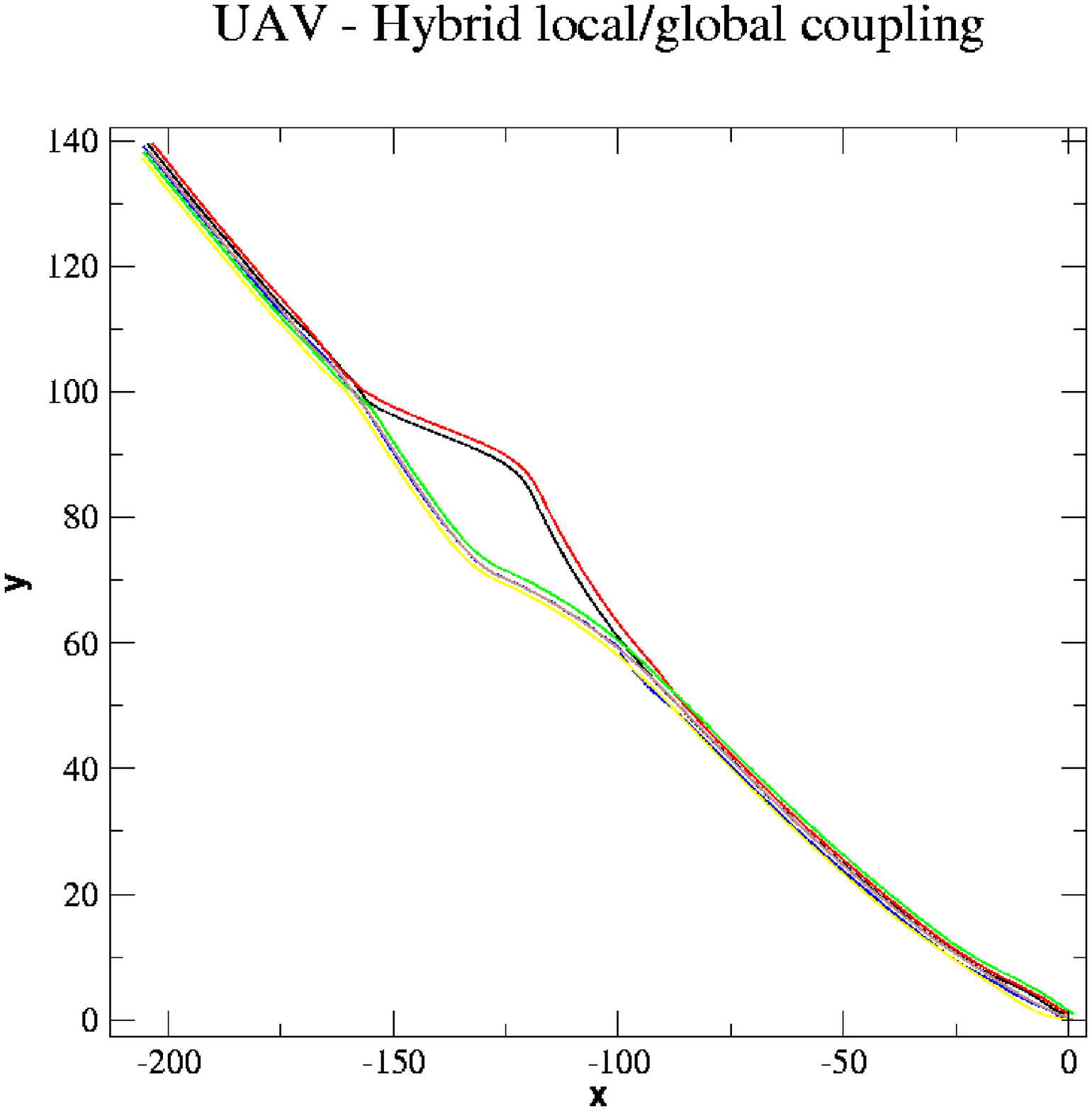}\end{center}

\begin{center}\includegraphics[%
  scale=0.5]{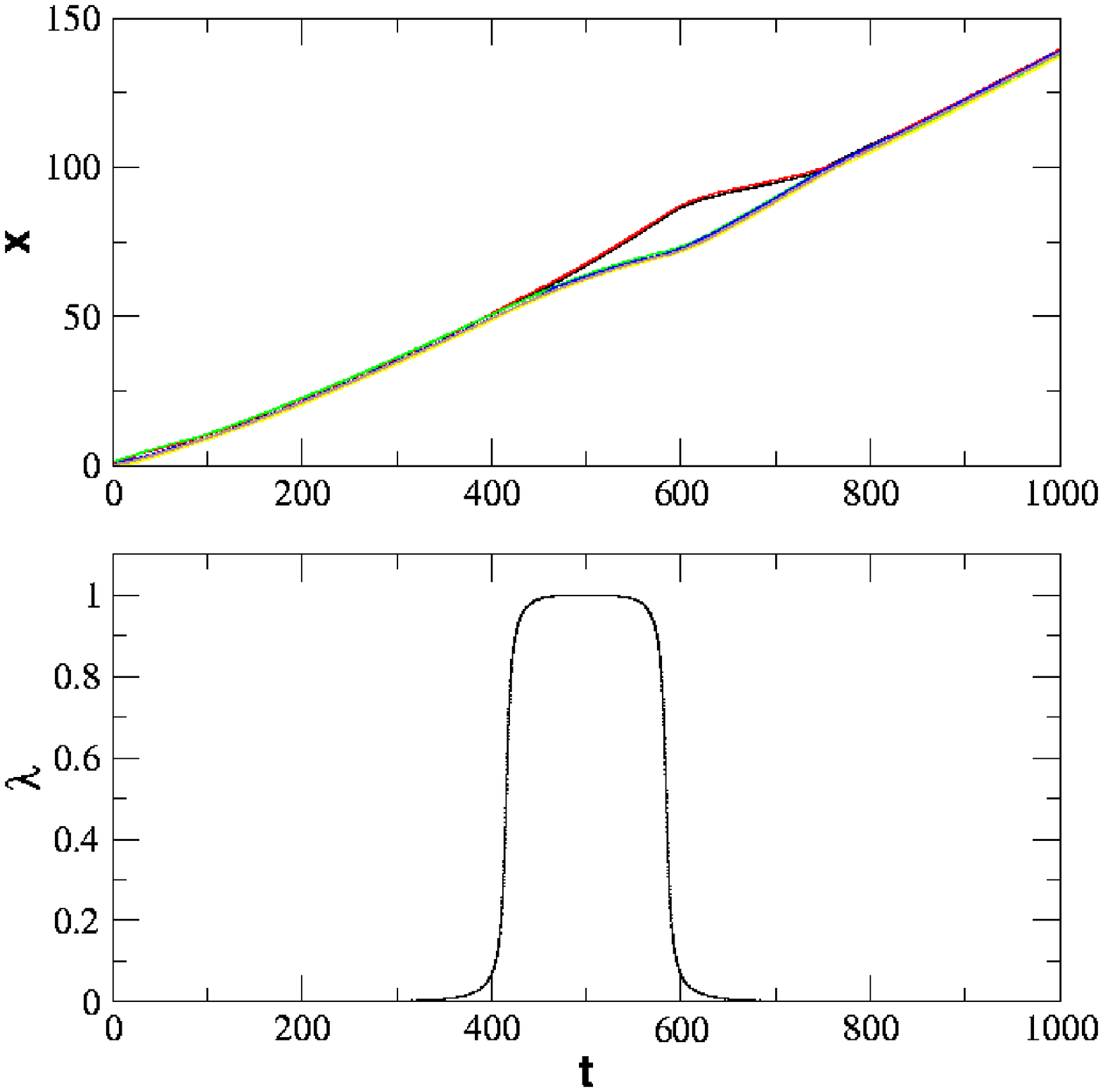}\end{center}

\caption{\label{cap:UAV_hybrid_coupling}A simulation of the UAV model using
the homotopy control law (\ref{eq:homotopy_ctrl_law}). The upper
figure show a time-trace of the simulation in the $\mathbf{r}=(x,y)$
plane, the center plot shows $x$ vs $t$, and the lower plot, $\lambda$
vs $t$. Initially, $\lambda=0$ and the global, group averaged control
law is used. Agents start out in arbitrary directions, but soon organize
into a unified, coherent swarm. Near $t=400$, the homotopy parameter
is turned on ($\lambda=1$), which switches the control law to local
coupling with leader following motion, and the swarm splits into two
subswarms. Near $t=600,$ the homotopy parameter is switched off and
global coupling is re-enabled, and a single swarm automatically reforms.}
\end{figure}
\end{center}

We first present a prototypical simulation using the homotopy control
law presented in section \ref{sec:Homotopy-control-law.}, using $n=6$
agents in the swarm. The parameters are $\alpha=\eta=\mu=0.02$ and
$r_{0}=1.5$, while the cutoff function parameters are $q=0$ and
$w=4$, and simulations commence with $\lambda=0$, so that the initial
control law is the globally coupled group averaged law (\ref{eq:control_law}).
The initial positions of the six agents are $\{(-1,1),(0,1),(1,1),(-1,0),(0,0),(1,0)\}$,
while the initial orientations $\theta_{i}$ of each agent is randomly
chosen from angles constrained to lie $\phi-\pi/4<\theta_{i}<\phi+\pi/4$,
and $\phi\in[0,2\pi]$ is also randomly chosen, and the system is
integrated to $t=1000$. Near $t=400,$ the homotopy parameter $\lambda$
increases to one, and the control laws (\ref{eq:follower_control})
and (\ref{eq:leader_control}) are used. See figure \ref{cap:UAV_hybrid_coupling}.
As the homotopy parameter is switched on, the two leaders use the
simple 'programs' $s_{1}=0.01$ and $s_{2}=-0.01$. This causes one
leader to start a gentle clockwise loop and the other leader to begin
a gentle counterclockwise loop. At this point in the simulation, the
entire swarm then splits into two subswarms as the follower agents
move toward their nearest leader agent. As the homotopy parameter
is decreased to zero near $t=600,$ the control law reverts to (\ref{eq:control_law}),
global communication between all agents is restored, and the two subswarms
reform as one swarm. 

\begin{center}%
\begin{figure}
\begin{center}\includegraphics[%
  scale=0.5]{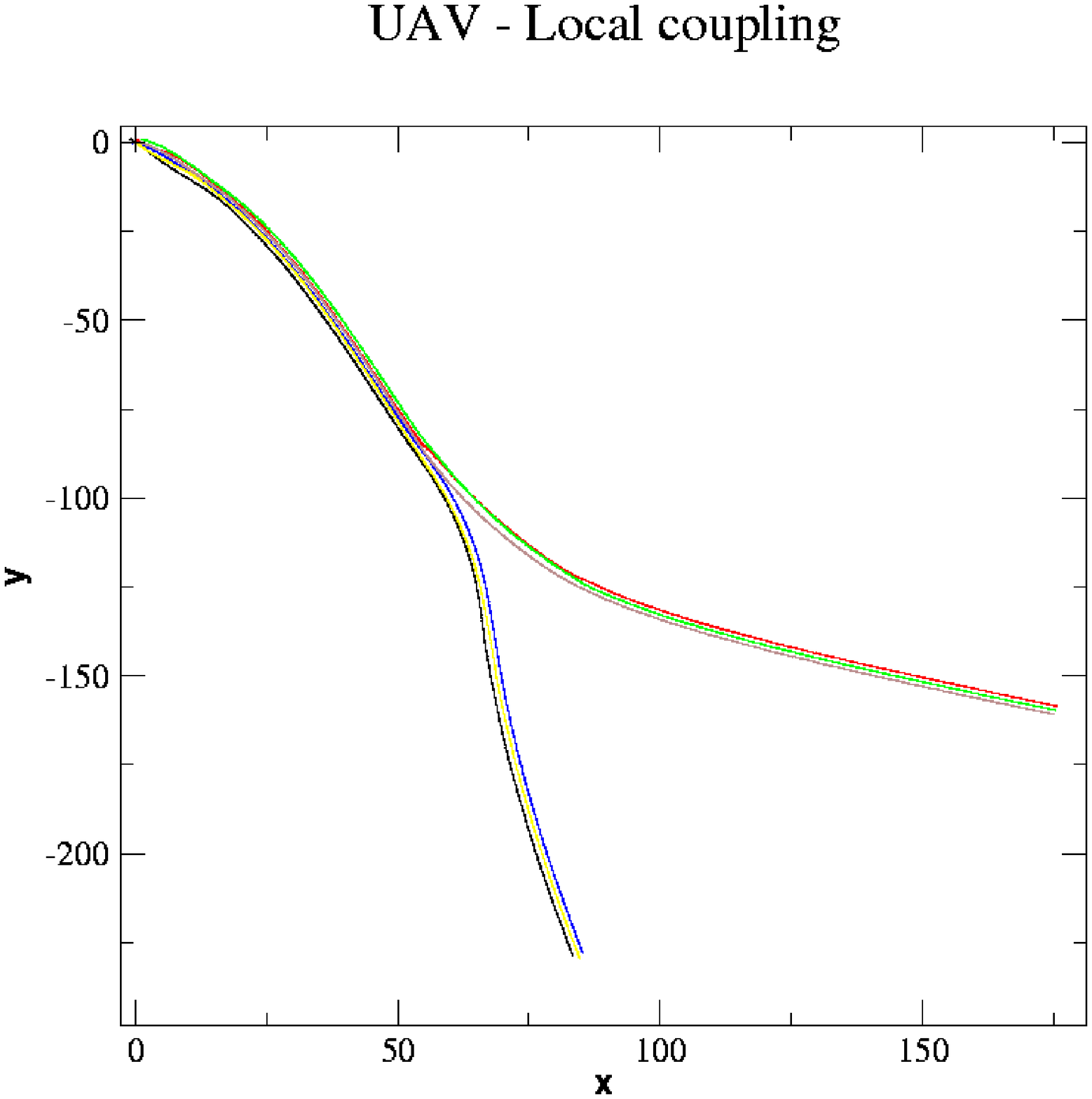}\end{center}

\begin{center}\includegraphics[%
  scale=0.5]{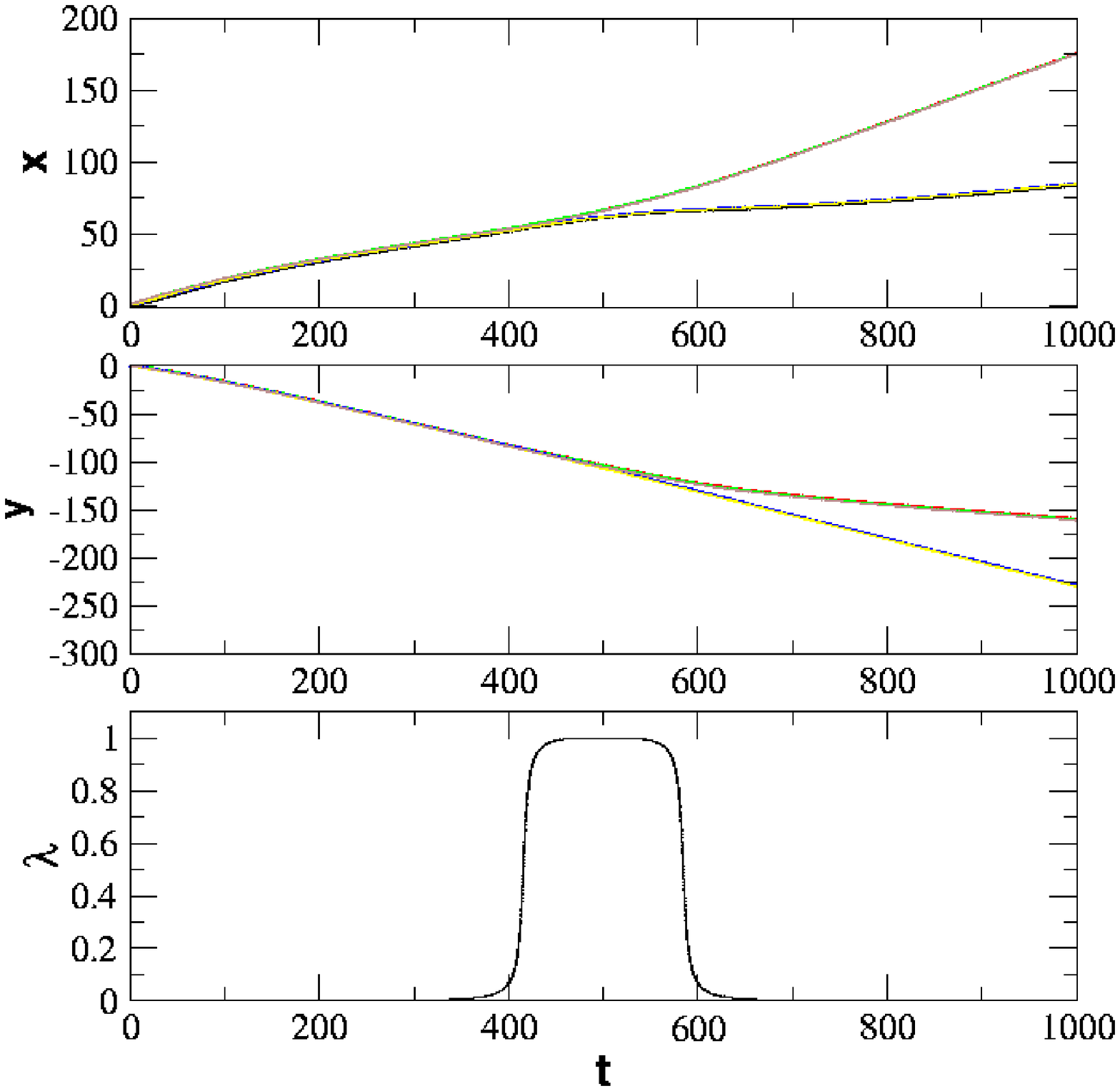}\end{center}

\caption{\label{cap:UAV_local_coupling}A simulation of the UAV model using
local coupling only. The upper figure show a time-trace result of
the simulation in the plane, while the lower plot shows $r_{x}$ vs
$t$, $r_{y}$ vs $t$ and $\lambda$ vs $t$, respectively. Initially,
the full swarm is in communication, $\lambda=0$ and the locally coupled,
group averaged control law is used. As in Fig. \ref{cap:UAV_hybrid_coupling},
agents soon organize into a unified, coherent swarm. Near $t=400$,
the homotopy parameter is turned on ($\lambda=1$), which switches
the control law to leader following motion, and the swarm splits into
two subswarms. Near $t=600,$ the homotopy parameter is switched off
and group averaged motion is re-enabled. Due to the fact that local
coupling is used and the subswarm clusters are far apart, a single
swarm does not reform in this case.}
\end{figure}
\end{center}

We next present an extension of the above, a prototypical simulation
using the homotopy control law, but with local coupling throughout,
so that the simulation commences with locally coupled, group averaged
motion, with all agents in communications range of one another. As
in the previous subsection, the homotopy parameter is initially zero,
but near $t=400$ it is increased to one, and the system transitions
from group averaged motion to leader following motion, with two leaders.
Once again the swarm splits into two subswarm clusters for the same
reasons as outlined in the previous subsection. Near $t=600,$ the
homotopy parameter is then decreased to zero, and the system returns
to group averaged motion. See figure \ref{cap:UAV_local_coupling}.
However, due to the local nature of the coupling, once the system
returns to group averaged motion, the two subswarms remain independent,
since the subswarms are not in communications range of one another. 

\section{Target seeking and barrier avoidance control laws\label{sec:Target-seeking-laws}}

The homotopy control law of the preceding section provided a mechanism
to switch between global and local coupling, and thus created a swarm
which could split into subswarms, and each of the subswarms could
be independently directed. We now introduce an additional method for
both swarm navigation, by coding an \emph{a priori} target-seeking
behavior into the control law, and a method of barrier avoidance,
by automatically sensing a barrier and splitting the swarm around
the barrier. We note that control laws presented in this section may
also be integrated with the previously introduced control laws. The
character of the ideas in this section is similar in spirit to those
presented in \cite{ChangShaddenMarsden}, where a force law is introduced
with a global potential for target seeking, but where both gyroscopic
and braking forces are used for collision avoidance.

\subsection{Target seeking control law.}

A target is considered here to be a fixed point in the plane that
is specified ahead of time. In the context of UAVs, this implies that
the position of a target would be preprogrammed before deployment,
though an alternate possibility would be to communicate target coordinates
to agents in flight. We introduce a target in the model by globally
coupling the agents to an 'agent' that does not move. Let $\bar{r}$
be the fixed location of the target. The modified control law for
target seeking is\begin{eqnarray}
u_{k} & = & \sum_{j\neq k}\left[c(|r_{jk}|,0,w)(-\alpha\left(1-\left(\frac{r_{0}}{\left|r_{jk}\right|}\right)^{2}\right)(r_{jk}\cdot y_{k})\right]+\label{eq:modif_ctrl_law_2}\\
 &  & \gamma\alpha\left(1-\left(\frac{r_{0}}{\left|\bar{r}_{k}\right|}\right)^{2}\right)(\bar{r}_{k}\cdot y_{k}).\nonumber \end{eqnarray}
where $\bar{r}_{k}$ is the vector directed from the position of the
$k^{\textrm{th}}$ agent to $\bar{r}$, and $\gamma$ is a weighting
constant. Note that there is no term in this control law to align
agents to a common heading. In fact, the only inter-agent term is
the first term, involving the summation, which provides for collision
avoidance. The cutoff function $c$ implies that this term will have
no effect if agents are outside of the cutoff radius $w$. The second
term, which is global, steers individual agents toward the target.
Though there are no terms to explicitly group agents, if the initial
conditions are chosen so that the agents start in a group, then they
will tend to stay together, as they collectively steer toward the
target.

\subsection{Barrier avoidance}

The homotopy control law presented in section \ref{sec:Homotopy-control-law.}
can be used for barrier avoidance by splitting a swarm into two subswarms,
which are steered independently around the barrier. The swarm splitting
is achieved by explicitly changing the homotopy parameter $\lambda$.
In contrast, we consider here an additional term for the control law,
which applies an angular force to an agent when it is within sensing
range of a barrier, where the position of the barrier is \emph{a priori}
unknown. 

We restrict ourselves to the case of a convex and stationary barrier
in $\Re^{2}$. For our purposes, a barrier $B$ is the convex hull
defined by a set of $m$ points $b_{i}\in\Re^{2}$. The location of
the barrier is not \emph{a priori} known to an agent. Instead the
barrier is detected whenever an agent is within range of any of the
points defining the barrier, in which case we say that the agent is
within range of the barrier. 

Barrier avoidance logic is implemented in the model as follows. For
each agent, at every time step of the simulation, we calculate a vector,
the \emph{average barrier direction vector}, and which may be the
zero vector if the agent is not within a neighborhood of a barrier.
The average barrier direction vector (directed from the $k^{\textrm{th}}$
agent) is defined as,\begin{equation}
v_{k}=\left\{ \begin{array}{cc}
\frac{\sum_{i=1}^{m}\left[(b_{i}-y_{k})c\left(\left|b_{i}-y_{k}\right|,0,w\right)\right]}{\left|\sum_{i=1}^{m}\left[(b_{i}-y_{k})c\left(\left|b_{i}-y_{k}\right|,0,w\right)\right]\right|}, & \qquad\textrm{if }\left|\sum_{i=1}^{m}\left[(b_{i}-y_{k})c\left(\left|b_{i}-y_{k}\right|,0,w\right)\right]\right|\neq0\\
0, & \textrm{otherwise}\end{array}\right.,\label{eq:barrier_avodiance_law}\end{equation}
 where $c(\cdot)$ is the cutoff function (\ref{eq:cutfn}). The control
law is then modified by adding the term $\pm(v_{k}\cdot y_{k})s$,
where the sign is chosen to be the sign of the expression $v_{k}\cdot y_{k}^{\perp}$.
This term serves to steer the agent perpendicular to the direction
of the average vector $v_{k}$, and the sign is chosen to steer the
agent away from the the average direction of the barrier, relative
to the current heading of the vehicle. This can result in a splitting
of the swarm into two subswarms, with one subswarm going around one
side of the barrier, and the other subswarm going around the other
side.

\subsection{Simulation results for target-seeking and barrier avoidance control
law}

\begin{figure}
\begin{center}\includegraphics[%
  scale=0.35]{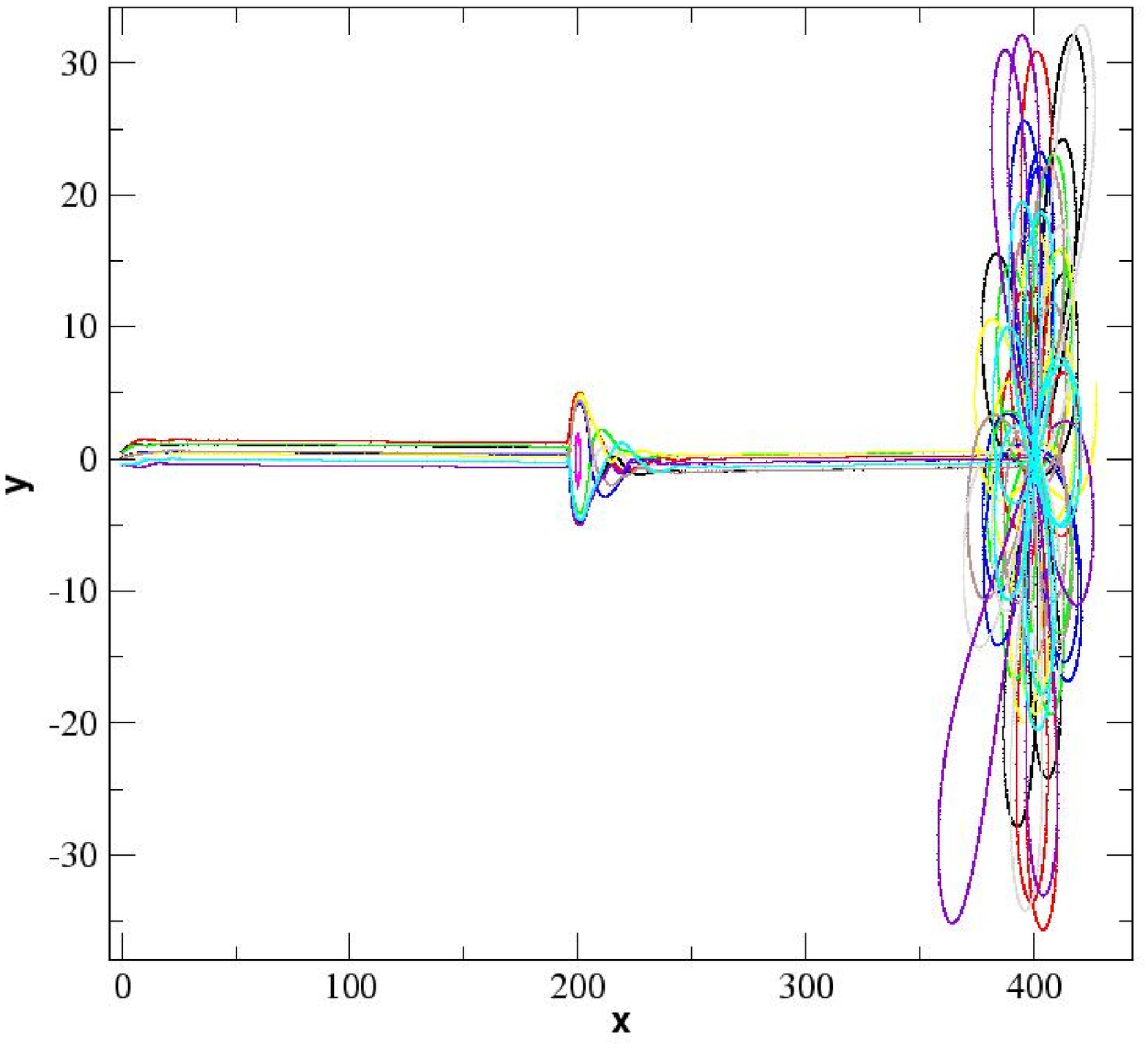}\includegraphics[%
  scale=0.35]{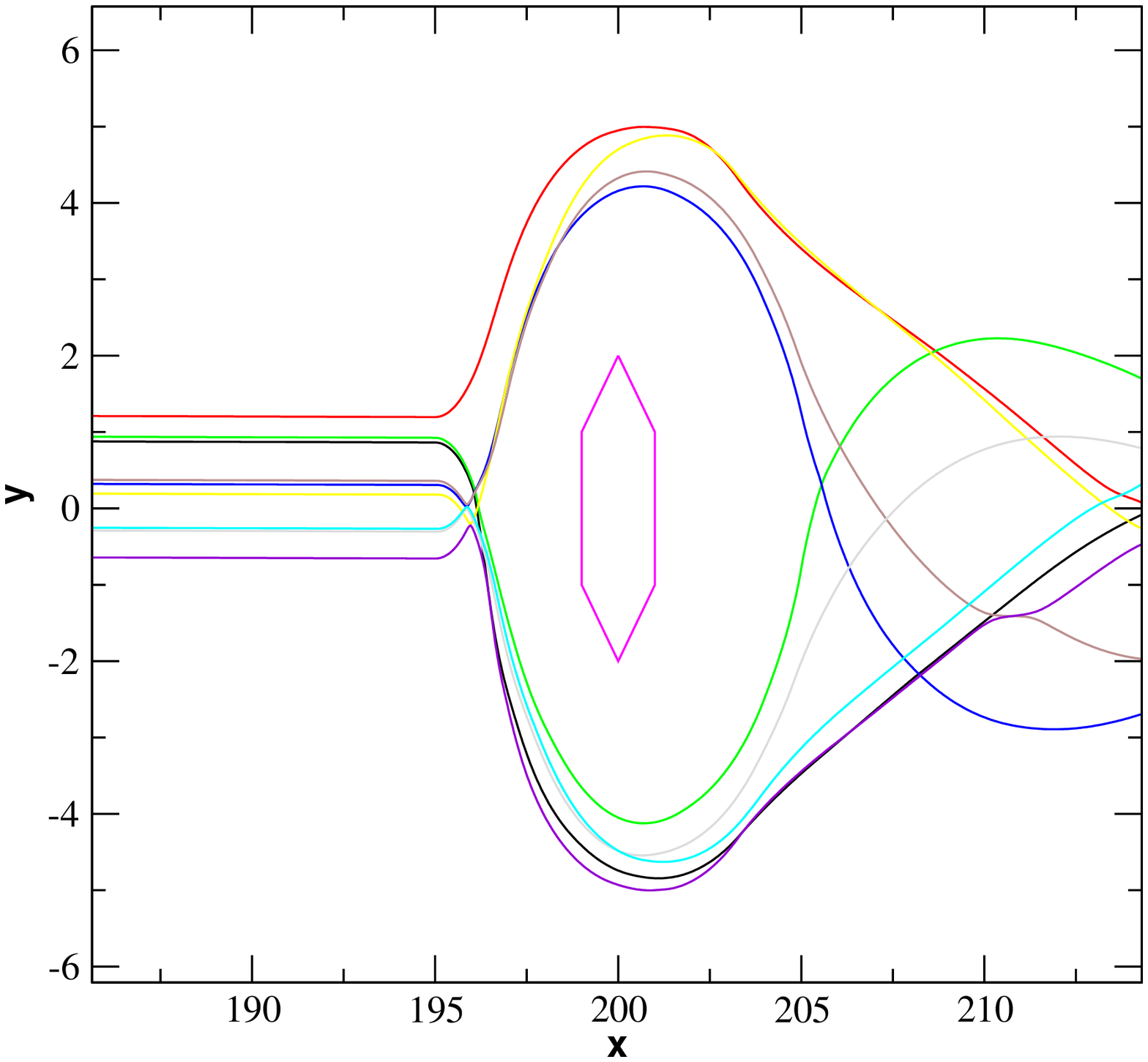}\end{center}

\caption{\label{cap:Target_seeking_2}Results of a simulation using the control
law of equation (\ref{eq:modif_ctrl_law_2}). The figure on the left
shows a time-trace of the complete simulation. The swarm travels straight
to the target point, which is at $(x,y)=(400,0)$, while avoiding
the barrier centered at $(x,y)=(200,0)$. The figure on the right
shows a detail of the barrier avoidance.}
\end{figure}

We present the results of a simulation of the model using the control
law \ref{eq:modif_ctrl_law_2} with the barrier avoidance control
\ref{eq:barrier_avodiance_law}. Figure \ref{cap:Target_seeking_2}
shows the result of a typical simulation. A target point is located
at $(x,y)=(400,0)$, and there is a hexagonal barrier centered at
$(x,y)=(200,0)$, and is defined by the points $\{(200,2),(201,1),(201,-1),(200,-2),(199,-1),(199,1)\}$.
As the swarm approaches the barrier, it again splits into two subswarms.
All agents rejoin into a single swarm after the barrier is passed
and continue on to the target. This time, upon arrival, the agents
swarm in an irregular fashion about the target point. The irregular
swarming about the target is due to the lack of inter-agent baseline
controls implemented in this version of the control law.

\section{Discussion}

The presented Frenet-Serret model and associated control laws exhibits
robust and spontaneous coherent motion of a collection of $n$ agents
with controlled clustering for any smooth dynamical system. Such emergent
behavior is important in obstacle and predator avoidance. Cluster
formation from a coherent structure was done via a new type of control,
which we introduced as a homotopy control. Using a simple central
parameter, homotopy control provides an easily implementable method
to create new emergent behavior from coherent structures. The model
is robust in the sense that small perturbations of constituent agents
of the swarm results in little or no change in the coherent motion
of the swarm as a whole. We tested this by introducing additive noise
into the simulations. At each time step, the positions and angles
of the agents were perturbed with a small amount of noise. Noise was
taken from a uniform distribution with mean zero. Results of simulations
were qualitatively similar to those presented in section \ref{sec:Homotopy-control-law.}.

On the other hand, we also showed in section \ref{sec:Target-seeking-laws}
that even with very loose coupling, involving only simultaneous target
seeking, and where the only inter-agent coupling is via a collision
avoidance term, ordered behavior can emerge, even when obstacle avoidance
is taken into account.

Previous studies have focused on presenting unified coherent motion
of a swarm. We have extended these results by presenting a method
to automatically transition to subswarm clusters, formed from an original
larger swarm, and functioning independently. The spontaneous coherence
implies that individual agents do not need to be manually controlled.
Indeed, that is one main goal of such research; to find a set of (preferably
simple) rules which will result in the desired behavior with a high
degree of autonomy for the swarm, and with a minimum of external inputs.

There are some limitations to the currently considered model. When
using only local coupling with the homotopy control law, there is
no way to reunite subswarm clusters, and a separate mechanism would
have to be introduced to do so. Additionally, the leader following
model presented is asymmetric, in that there is a distinction between
leader and follower agents. Thus, if a leader agent is disabled, the
subswarm cluster is no longer controllable. A better approach would
be to consider a symmetric control law that doesn't distinguish between
leader and follower agents, but which maintains similar behaviors.
This is the subject of ongoing research. 

One obvious extension of the current model is to obtain a dictionary
of useful controls which can be strung together in a similar fashion
to what we have done with the homotopy control law, perhaps with multiple
homotopy parameters, in order to obtain multiple emergent behaviors.
Additionally, a stochastic control law, in which the swarm maintains
a loose cohesiveness, while incorporating stochastic motion to avoid
interception by predators, is also being explored as an extension.

\begin{ack}
We thank E Justh and P.S. Krishnaprasad for useful conversations which
lead to the new control scheme presented.
\end{ack}
\bibliographystyle{/home/dmorgan/work/Swarming/UAV_writeup/Working/elsart-num}

\begin{thebibliography}{10}
\expandafter\ifx\csname url\endcsname\relax
  \def\url#1{\texttt{#1}}\fi
\expandafter\ifx\csname urlprefix\endcsname\relax\def\urlprefix{URL }\fi

\bibitem{JK04}
E.~W. Justh, P.~S. Krishnaprasad, Equilibria and steering laws for planar
  formations, Systems and Control Letters 52 (2004) 25--38.

\bibitem{Bonabeua99}
D.~M. Bonabeua, H., G.~Theraulaz, Swarm Intelligence: From Natural to
  Artificial Systems, Oxford Univ. Press, 1999.

\bibitem{JK01}
E.~W. Justh, P.~S. Krishnaprasad, Pattern-forming systems for control of large
  arrays of actuators, J. Nonlinear Sci. 11 (2001) 239--277.

\bibitem{EK01}
L.~Edelstein-Keshet, Mathematical models of swarming and social aggregation,
  invited lecture; The 2001 International Symposium on Nonlinear Theory and its
  Applications, (NOLTA 2001) Miyagi, Japan.

\bibitem{Murray}
J.~D. Murray, Mathematical Biology, Springer-Verlag, Berlin, 1993, second
  edition.

\bibitem{EKWG98}
L.~Edelstein-Keshet, J.~Watmough, D.~{Gr\"unbaum}, Do travelling band solutions
  describe cohesive swarms? {A}n investigation for migratory locusts, J. Math.
  Biol. 36~(6) (1998) 515--549.

\bibitem{MEK99}
A.~Mogilner, L.~Edelstein-Keshet, A non-local model for a swarm, J. Math. Biol.
  38~(6) (1999) 534--570.

\bibitem{WEK95}
J.~Watmough, L.~Edelstein-Keshet, A one-dimensional model of trail propagation
  by army ants, J. Math. Biol. 33~(5) (1995) 459--476.

\bibitem{SET01}
F.~Schweitzer, W.~Ebeling, B.~Tilch, Statistical mechanics of
  canonical-dissipative systems and applications to swarm dynamics, Phys. Rev.
  E 64.

\bibitem{G99}
D.~Gr{\"u}nbaum, Advection-diffusion equations for generalized tactic searching
  behaviors, J. Math. Biol. 38~(2) (1999) 169--194.

\bibitem{MEK96}
A.~Mogilner, L.~Edelstein-Keshet, Spatio-angular order in populations of
  self-aligning objects: formation of oriented patches, Physica D 89~(3-4)
  (1996) 346--367.

\bibitem{Stocker}
S.~St{\"o}cker, Models for tuna school formation, Math. Biosciences 156 (1999)
  167--190.

\bibitem{Albano}
E.~V. Albano, Self-organized collective displacements of self-driven
  individuals, Phys. Rev. Lett. 77~(10) (1996) 2129--2132.

\bibitem{BDG97}
H.~J. Bussemaker, A.~Deutsch, E.~Geigant, Mean-field analysis of a dynamical
  phase transition in a cellular automaton model for collective motion, Phys.
  Rev. Lett. 78~(26) (1997) 5018--5021.

\bibitem{CBV99}
A.~Czir\'ok, A.-L. Barab\'asi, T.~Vicsek, Collective motion of self-propelled
  particles: Kinetic phase transition in one dimension, Phys. Rev. Lett 82~(1)
  (1999) 209--212.

\bibitem{VCFH99}
T.~Vicsek, A.~Czir\'ok, I.~J. Farkas, D.~Helbing, Application of statistical
  mechanics to collective motion in biology, Physica A 274 (1999) 182--189.

\bibitem{TT95}
J.~Toner, Y.~Tu, Longe-range order in a two-dimensional dynamical {XY} model:
  how birds fly together, Phys. Rev. Lett. 75~(23) (1995) 4326--4329.

\bibitem{TT98}
J.~Toner, Y.~Tu, Flocks, herds, and schools; a quantitative theory of flocking,
  Phys. Rev. E 58~(4) (1998) 4828--4858.

\bibitem{Tu00}
Y.~Tu, Phases and phase transitions in flocking systems, Physica A 281 (2000)
  30--40.

\bibitem{Irving:1950}
J.~H. Irving, J.~G. Kirkwood, The statistical theory of transport processes.
  {IV}. {T}he equations of hydrodynamics, Journal of Chemical Physics 18~(6)
  (1950) 817--829.

\bibitem{Thompson:1972}
C.~J. Thompson, Mathematical Statistical Mechanics, Princeton University Press,
  1972.

\bibitem{Mazo:1967}
R.~M. Mazo, Statistical mechanical theories of transport processes, Vol.~1 of
  The International Encyclopedia of Physical Chemistry and Chemical Physics,
  Pergamon Press, 1967.

\bibitem{ChangShaddenMarsden}
D.~E. Chang, S.~C. Shadden, J.~E. Marsden, R.~Olfati-Saber, Collision avodiance
  for multiple agent systems, Proc. of the 42nd IEEE Conf. on Decision and
  Control.

\bibitem{zohdi03}
T.~I. Zohdi, Computational design of swarms, Int. J. Num. Meth. Eng. 57 (2003)
  2205--2219.

\bibitem{Allgower80}
E.~L. Allgower, K.~Georg, Simplicial and continuation methods for approximating
  fixed points and solutions to systems of equations, SIAM Review 22 (1980)
  28--96.

\bibitem{Rheinboldt00}
J.~M. Ortega, W.~C. Rheinboldt, Iterative solution of nonlinear equations in
  several variables, Society for Industrial and Applied mathematics, 2000.

\bibitem{Jurdjevic}
V.~Jurdjevic, Geometric Control Theory, Cambridge Univ. Press, 1997.

\bibitem{JK02}
E.~W. Justh, P.~S. Krishnaprasad, A simple control law for uav formation
  flying, ISR Technical Report 2002-38.

\bibitem{JK03}
E.~W. Justh, P.~S. Krishnaprasad, Steering laws and continuum models for planar
  formations, Proc. IEEE Conf. Decision and Control (to appear).

\end{thebibliography}

\end{document}